\documentclass[a4paper,fleqn]{cas-dc}

\usepackage[numbers]{natbib}

\def\tsc#1{\csdef{#1}{\textsc{\lowercase{#1}}\xspace}}
\tsc{WGM}
\tsc{QE}

\begin{document}
\let\WriteBookmarks\relax
\def\floatpagepagefraction{1}
\def\textpagefraction{.001}
\let\printorcid\relax 

\shorttitle{Design of a Noval Wearable ECG Monitoring Device}    

\shortauthors{Zhengbao Yang et al.}

\title[mode = title]{Design of a Noval Wearable ECG Monitoring Device}

\author[1]{Ruihua Wang}

\author[1]{Mingtong Chen}

\author[1]{Zhengbao Yang}
\cormark[1] 
\ead{zbyang@hk.ust} 
\ead[URL]{https://yanglab.hkust.edu.hk/}

\address[1]{The Hong Kong University of Science and Technology
Hong Kong, SAR 999077, China}

\cortext[1]{Corresponding author} 

\begin{abstract}
The aim of this project is to develop a new wireless powered wearable ECG monitoring device. The main goal of the project is to provide a wireless, small-sized ECG monitoring device that can be worn for a long period of time by the monitored person.
Electrocardiogram ECG reflects physiological and pathological information about heart activity and is commonly used to diagnose heart disease. Existing wearable smart ECG solutions suffer from high power consumption in both ECG diagnosis and transmission for high accuracy. Monitoring of ECG devices is mainly done by data extraction and acquisition, pre-processing, feature extraction, processing and analysis, visualisation and auxiliary procedures. During the pre-processing of the information, different kinds of noise generated during the signal collection need to be taken into account. The quality of the signal-to-noise ratio can usually be improved by optimising algorithms and reducing the noise power. The choice of electrodes usually has a direct impact on the signal-to-noise ratio and the user experience, and conventional Ag/AgCl gel electrodes are not suitable for long-term and dynamic monitoring as they are prone to skin irritation, inflammation and allergic reactions. Therefore, a completely new way of combining electrodes and wires will be used in the report. The electrodes and wires are cut in one piece from a silver-plated fabric. The wire portion is cut into a curved structure close to an ‘S’ shape to ensure that it has good ductility for comfort and signal integrity during daily movement of the garment. In addition, a conductive gel can be used to enhance the signal collection at the electrode's contact position and, being non-allergenic and easy to clean, it can be used to enhance the signal in some special situations.
This research improves the reliability, convenience and sustainability of wearable ECG monitoring devices in the daily use of the user by optimising the energy supply system and the information pre-processing part. In the future, more portable and long-lasting wearable devices will be introduced through deeper research on energy harvesting.

\end{abstract}




\maketitle

\section{Introduction}

According to data from the World Health Organization (WHO), cardiovascular diseases are the leading cause of death globally, resulting in 17.9 million deaths each year \cite{1}. Driven by health awareness, more people are willing to self-monitor their health, which has led to the widespread adoption of wearable devices. There is also an urgent need for lightweight and low-power long-term monitoring instruments in clinical applications \cite{2}. Therefore, continuous heart rate monitoring and instant heartbeat detection are major focuses of contemporary healthcare. Electrocardiograms (ECG) can reflect physiological and pathological information about heart activity and are commonly used for diagnosing heart disease.

Existing wearable smart ECG solutions face the problem of excessive power consumption during ECG diagnosis and high-precision transmission. The monitoring of ECG devices mainly involves data extraction and collection, preprocessing, feature extraction, processing and analysis, visualization, and auxiliary programs \cite{3}. Detailed or accurate ECG signals often require longer time scales (i.e., several days) for recording. As a result, the generated records produce huge amounts of data, which need to be stored in storage space or transmitted over networks \cite{4}. For example, Yishan Wang et al. developed a small wireless ECG sensor node, along with a ZigBee coordinator and a graphical display interface. This system achieved power consumption reductions of 20\% and 30\% during normal activity and rest periods, respectively, by utilizing dynamic adjustments of received signal strength indicators (RSSI) and power levels \cite{2}. However, the above solutions are still only in the laboratory stage and have not been put into the market, so the functional approach used in this paper is still mainly powered by rechargeable lithium batteries.

In many cases, ECG is recorded under moving or strenuous conditions, which leads to interference from various types of noise. Therefore, noise reduction is an important objective of ECG signal processing. For example, there is significant spectral overlap between ECG and muscle noise. However, when appropriately utilizing the ECG as a repetitive signal, muscle noise can be reduced \cite{4}. To improve the signal-to-noise ratio (SNR), different algorithms can be used to optimize results for various types of noise. For instance, the Equiripple notch filter is the best choice for eliminating power line interference, while discrete Meyer wavelet and improved threshold functions should be used to eliminate motion artifacts and EMG noise, or empirical mode decomposition can be applied for baseline wander \cite{5}. The choice of electrode also affects both the quality of the signal-to-noise ratio and the experience of the monitored individual. Traditional Ag/AgCl gel electrodes, the most common type, have been shown to be unsuitable for long-term and dynamic monitoring as they can easily cause skin irritation, inflammation, and allergic reactions \cite{2}. Lei et al. manufactured a self-adhesive, stretchable, and conformable epidermal dry electrode based on acid-modified silk fibroin (AM-SF)/cellulose nanocrystal (CNC) films, a Ppy conductive layer, and a calcium-modified SF adhesive layer, capable of reliably recording ECG and EMG signals. Moreover, this electrode can be easily removed from the skin with simple rinsing to reduce harm to the body, and it possesses excellent enzymatic degradability, allowing for easy degradation without leaving harmful substances \cite{6}.

Therefore, as a product targeting commercial products, most of the materials used in this paper are mature commercial products. The commercialisation of some components can provide good realisability and production guarantee for the birth of the product, and reduce the technical obstacles from R\&D to mass production. For example, the selected core components, such as conductive fabrics, conductive gels, electrode structures, and power supply systems, are all derived from existing technology platforms or proven commercial solutions. This strategy not only reduces R\&D costs and time, but also helps to ensure the stability and user acceptance of the device in both clinical and civilian environments. In addition, prioritising mature components that can be mass-produced also facilitates the large-scale rollout and approval of future devices. By balancing innovative design with realisability, this project seeks to strike a good balance between scientific research and market translation.

\section{Design and methodology}
\begin{figure}[h]
	\centering
		\includegraphics[scale=1]{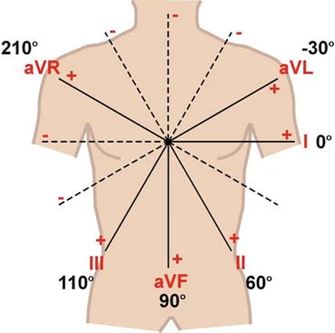}
	  \caption{Cardio axis}\label{FIG:1}
\end{figure}
Twelve-lead electrocardiogram is considered as the ‘gold standard’ for clinical electrocardiographic monitoring. It is constructed by using three electrodes on the limbs: right arm RA, left arm LA, left leg LL, right leg RL (of which RL serves as the ground lead for correcting the electrical signals), and six electrodes on the chest (V1-V6), thus providing a 360-degree view of the electrical activity in the orthogonal coronal and horizontal planes. and six chest leads, thus providing a 360-degree view of the electrical activity of the heart in the orthogonal coronal and horizontal planes. The limb leads were divided into standard leads I, II, and III and augmentation leads aVR, aVL, and aVF, which depicted the electrical vectorial changes of the heart in the coronary plane by using the potential difference between the different electrodes or the reference mean potential; where the right upper extremity electrode was the negative electrode and the left upper extremity electrode was the positive electrode. The flow of electrical activity from the negative pole to the positive pole corresponds to the recording of electrical activity from the lateral angle of the left ventricle in lead I, resulting in a lead I ECG. Similarly, the ECG in leads II and III is derived \cite{7}. These three leads form the famous Einthoven's triangle, an approximately equilateral triangle, with the heart right in the centre of the Einthoven's triangle. For the three unipolar enhancement leads aVR, aVL, and aVF, an ECG in the aVR lead is derived by using two leads as electrodes, e.g., the right upper extremity as the positive pole, and the potential difference between the left upper and left lower extremities as the negative pole. Similarly, the ECG of aVL and aVF can be obtained by the same method. Finally, the ECG axial map can be obtained by combining the six leads together.

A positive waveform is produced when the cardiac vectors are in the same direction as the flow of electrical activity, a negative waveform is produced when the opposite is true, and an equipotential waveform is formed if the two are perpendicular. Because the direction of the lead II observation is essentially the same as the direction of the cardiac vectors, the standard PQRST waveform for the ideal situation is derived.
\begin{figure}[h]
	\centering
		\includegraphics[scale=1]{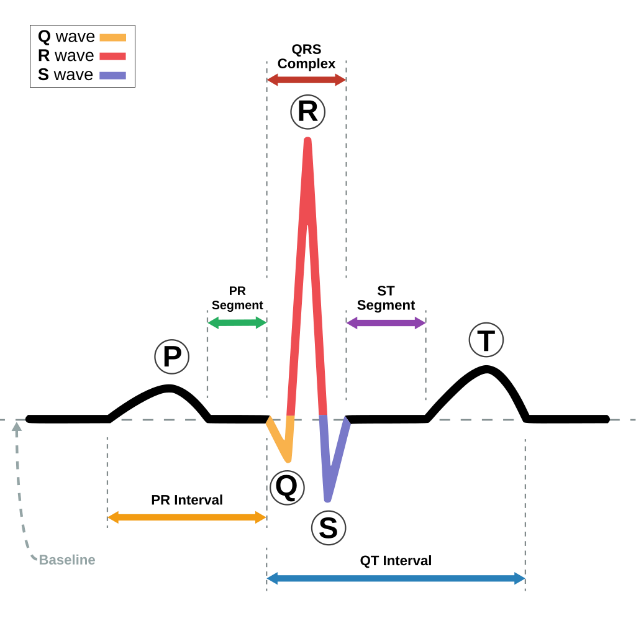}
	  \caption{The ideal standard PQRST waveform}\label{FIG:2}
\end{figure}
The limb leads described above are all electrical conduction vectors in the coronary plane of the heart. However, the heart is a three-dimensional structure, and one coronary plane is not enough, so there are also six electrode sheets affixed to the chest to form six unipolar thoracic leads to monitor the electrical conduction activity of the heart from the horizontal plane. The thoracic leads, on the other hand, were arranged sequentially along the right edge of the sternum to the left axillary midline, recording the propagation of electrical activity from right to left and back to front in the horizontal plane as the electrodes from V1 to V6 gradually moved towards the free wall of the left ventricle. The heart acts as the negative electrode V1-V6 act as the positive electrode and they produce different waveforms. With the help of these twelve synchronised leads, clinicians are able to accurately measure, for example, heart rate and rhythm, and to determine a wide range of pathological conditions such as arrhythmia.
\begin{figure*}[h]
	\centering
		\includegraphics[scale=1]{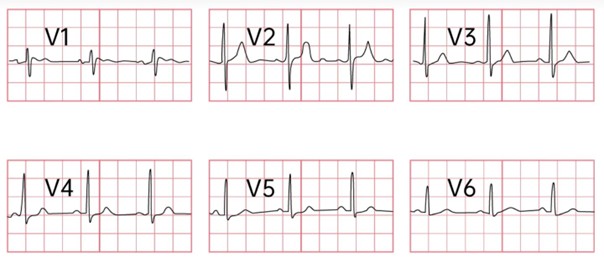}
	  \caption{Comparison of waveforms V1-V6}\label{FIG:3}
\end{figure*}
Each moment of electrical activity in the heart consists of many cardiomyocytes undergoing depolarisation and repolarisation at the same time, resulting in electrical vectors of different directions and sizes, which are superimposed to form a composite electrical vector, or composite vector for short. Thinking mathematically, when two vectors of the same direction are superimposed, their composite vector is the sum of these two vectors. If the two vectors are in opposite directions, their composite vector is the difference between the two. If two vectors are at an angle, their composite vector is the diagonal of their parallelogram. The magnitude and direction of the composite vectors are constantly changing over time, and each instant produces a composite vector that forms a loop curve by connecting the vertices of each composite vector, which is called a cardiac vector loop. Since the heart is a three-dimensional structure, it is also known as a spatial ECG vector loop. From atrial depolarisation, ventricular depolarisation and ventricular repolarisation, three types of spatial ECG vector loops can be obtained, namely the P loop, the QRS loop and the T loop. Taking the QRS ring as an example, it is projected vertically on the frontal, transverse and right side planes of the three-dimensional space, and the resulting projections are the planar ECG vectors. After completing the first projection, we continue to project the planar ECG vectors onto the lead axes of each lead. Take the frontal plane as an example, because the frontal plane is the coronal plane of the human body and corresponds to the six limb leads of the human body. In the case of lead I, for example, the direction of observation is from the positive zero degree of the lead axis, with the right half of the lead axis in the positive direction and the left half in the negative direction. When the cardiac activity starts, it passes through a very small negative potential before going through most of the remaining potentials. Therefore, the waveform of lead I is the result of a smaller negative waveform and a larger positive waveform. The same principle applies to the other leads, and all the waveforms finally make up the complete ECG waveform.

\begin{figure}[h]
	\centering
		\includegraphics[scale=1]{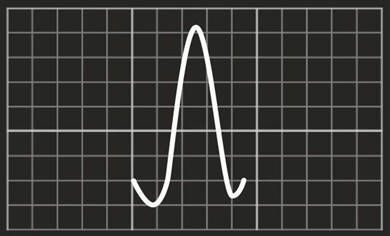}
	  \caption{QRS projections and waveform}\label{FIG:4}
\end{figure}
The quality of ECG is usually determined by a combination of factors such as the stability of the contact between the electrodes and the skin, the electrochemical properties of the electrode materials, the immunity of the wire transmission, and the amplification and filtering accuracy of the signal acquisition equipment. Among them, the quality of electrode contact is the first factor, if the contact between electrode and skin is poor, it will lead to baseline drift, IF interference or signal interruption. Secondly, the impedance characteristics of the electrode material affects the signal reduction. Conventional silver/silver chloride electrodes have excellent performance but rely on colloidal conductive paste, which makes them less comfortable to wear for a long period of time. The shielding capability and structural layout of the conductor also determine whether the entire signal path is susceptible to mechanical disturbances or electromagnetic noise.
Therefore, for the electrodes a rapid prototyping using 3D printing (Formlabs 3, flexible 80A material) was performed during the development phase, followed by uniform deposition of a biocompatible metal film on the surface of the microneedles by physical vapour deposition (PVD). The microneedles were tested for their density and quality due to their impact on the human body. Densities ranged from 1 to 46 needles, and from 1.5 to 69 needles, with ground-density electrodes being sharper but slightly more uncomfortable. In the tests it was found that the radius did not have a significant effect on impedance or comfort, but a height of over 1000 microns significantly affected comfort. The final choice was a radius of approximately 500 microns, a height of approximately 600 microns (0506), and a monolithic needle count of 46. The mechanical properties of the electrodes were tested against wet and dry electrodes. The impedance changes of the copper film and human skin were compared, and the results showed that the impedance level of the microneedle electrodes was between that of the wet and dry electrodes, i.e., it retained the advantage of the noise floor of the signal channel, and significantly improved the comfort and firmness of wearing.

\begin{table}[h]
\caption{Comparison of mechanical properties}\label{tbl1}
\begin{tabular*}{\tblwidth}{@{}LL@{}}
\toprule
 Against Copper Film Test & Against Human Skin Test \\ 
\midrule
 Wet (10$\Omega$) & Wet (550$\Omega$) \\
 Microneedle (1$\Omega$) & Microneedle (600$\Omega$) \\
 Dry (0.1$\Omega$) & Dry (700$\Omega$) \\
\bottomrule
\end{tabular*}
\end{table}
Keeping the wires from being damaged by stretching, perspiration or daily washing during wear is also necessary to ensure a high signal-to-noise ratio. For this reason, the electrodes are connected to wires made of conductive fibres, which are bonded and secured by a silver paste that is both adhesive and electrically conductive. The leads are made of commercially available conductive fibres that are commonly available on the market after cutting.
\begin{figure}[h]
	\centering
		\includegraphics[scale=0.75]{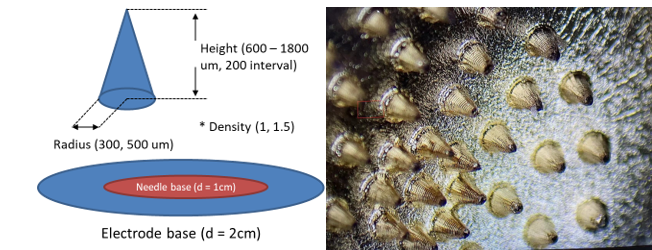}
	  \caption{Size of microneedle and the shape of microneedle after PVD under microscope}\label{FIG:5}
\end{figure}
\begin{figure*}[h]
	\centering
		\includegraphics[scale=0.8]{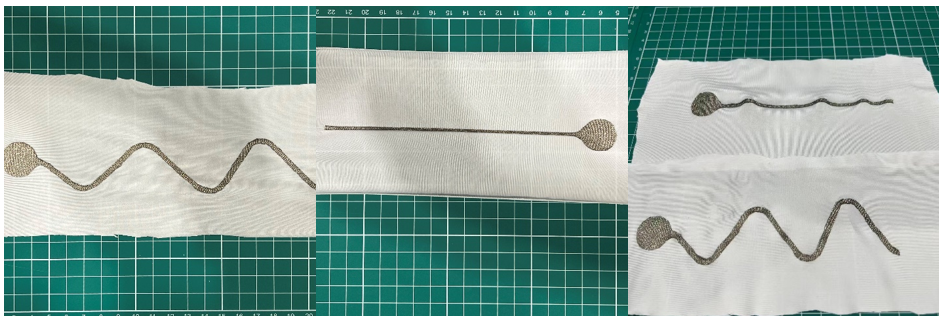}
	  \caption{Comparison of S-shape and linear shape}\label{FIG:6}
\end{figure*}
\begin{figure*}[h]
	\centering
		\includegraphics[scale=1]{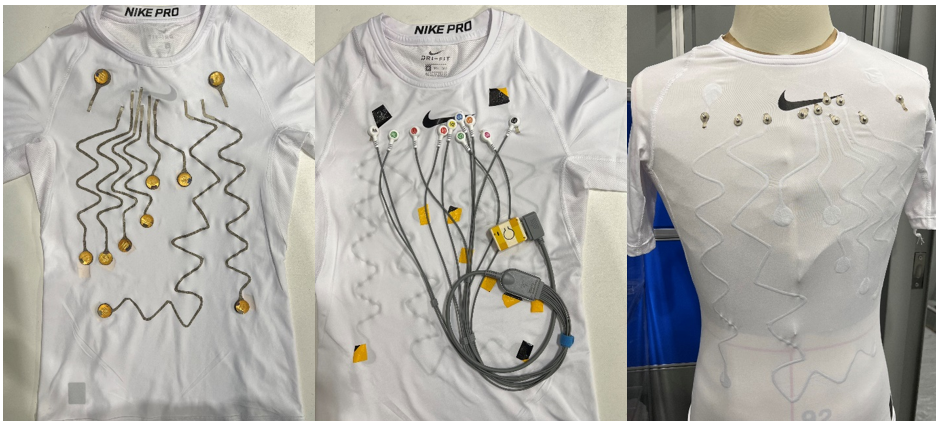}
	  \caption{Wearable ECG device demonstration}\label{FIG:7}
\end{figure*}

The wire path is designed according to the positional requirements of the twelve leads to be positioned, and the wire layout is designed as an S-shaped path. Compared to traditional straight wiring, the S-shaped structure demonstrates higher mechanical ductility and deformation cushioning during wear stretching or physical activity. In comparative tests, the resistance change rate of the S-shaped wire is significantly lower than that of a straight path of the same material after multiple stretching cycles. This flexible cushioning characteristic enables it to better adapt to local displacement, folding and pulling of the garment in dynamic use environments, thus effectively reducing the occurrence of broken wires, unstable signals or local desoldering. By optimising the bend radius, the S-shaped path maintains adequate electrical conductivity and improves the overall mechanical reliability of the system.

\begin{figure}[h]
	\centering
		\includegraphics[scale=1]{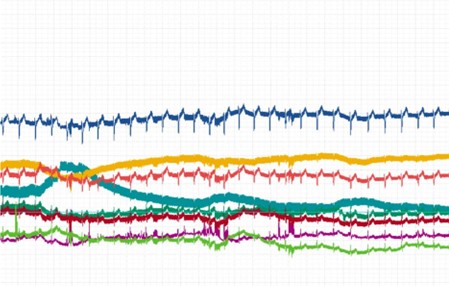}
	  \caption{Signal without electrodes}\label{FIG:8}
\end{figure}
\begin{figure}[h]
	\centering
		\includegraphics[scale=0.8]{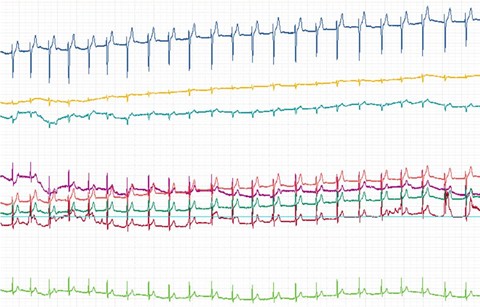}
	  \caption{Signal with electrodes}\label{FIG:9}
\end{figure}
\begin{figure*}[h]
	\centering
		\includegraphics[scale=1]{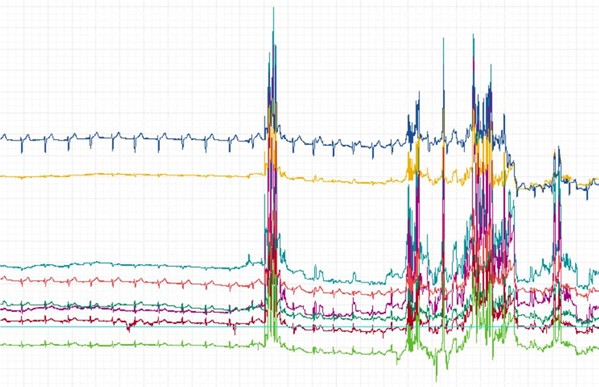}
	  \caption{The tester goes from a state of rest to a state of motion}\label{FIG:10}
\end{figure*}
The specific direction of the conductors was firstly drawn in CAD software, and the true body distance from the centre of the electrode patch to the sink port was tailored and calculated for each chest or limb conductor according to the aforementioned S-shaped layout principle; on this basis, a one-centimetre margin was reserved for each line to compensate for the dimensional changes of the fabric in bending and stretching, and the width of the line was uniformly set at 3 mm to take account of both the resistance and the flexibility. After completing the 2D graphics, the design file was fed into a CNC cutter, which was used to precisely cut the wire contours and positioning holes out of the entire sheet of PVC thermocompression film. The cut PVC film is then stacked with the uncut conductive fibre fabric in a positioning jig, and the film is fused to the fabric surface in a single pass through a 130 °C, 10-second hot press, which acts as an insulating layer and provides a clear reference for the cut edges during the subsequent cutting process. Once the film has cooled completely after hot pressing, the operator then cuts the wire profile with scissors along the edge of the film, maintaining a dimensional accuracy of ±0.2 mm thanks to the previously hardened edge of the film. Finally, the formed wire is placed in a predetermined position in the inner layer of the garment and pressed twice, at a lower pressure, with the same hot press, so that the PVC layer bites into the fabric fibres and is firmly bonded to the base fabric. At this point, a precise size, consistent direction, can be folded and washed with the clothing for many times without falling off the flexible wire, and its resistance is controlled within 0.05 $\Omega$/10 cm. According to the same process to prepare ten wires in turn, and in each wire at the end of the end of the garment to reserve a hole of about 3 mm in diameter. After hot pressing, a reinforcement ring made of the same material was added to the edge of the holes to prevent the fabric from tearing and expanding due to daily pulling; then the conductive fabric was folded out of the holes and connected to the stainless steel sub-buttons by riveting and pressing to realise the electro-mechanical integration. The outer side of the sub-buckle is reserved as a quick snap-in interface for the ECG acquisition harness, while the inner side forms a continuous low-resistance path with the conductive fabric, and at the same time serves as a stress cushion. The ten sub-buckles are arranged along the side of the garment according to the number of leads, so that the user only needs to insert the 12-lead ECG recorder's female harness at one time to complete the connection of all channels.

Thus, the electrical signals can be collected from the microneedle electrodes, and transmitted along the S-shaped flexible wire to keep intact during exercise, sweating and repeated washing, and finally enter the ECG acquisition unit, where the back-end circuitry completes the amplification, filtering, analogue-to-digital conversion, and wireless transmission to achieve the clinical-grade real-time ECG monitoring without affecting the wearing comfort and the durability of the clothing.

In order to verify the feasibility of this wearable system in real-life scenarios, signal tests were conducted when metal electrodes were not applicable and when metal electrodes were used.

The two screenshots above were taken on the same set of equipment, in the same subject, and with the same patch arrangement, and differ only in the point of whether or not a metal electrode sheet was used at the lead. The longitudinal comparison clearly shows the role of the metal electrode pads in reducing contact impedance, stabilising the baseline and suppressing motion, and polarisation artefacts.
In the screen without electrode pads, all chest leads (USERDATA 3-8) and limb leads (USERDATA 1 and 2) showed varying degrees of drift and segment dropout:The QRS wave was still recognisable, but with dense jagged high-frequency jitter attached to it; this type of jitter was mostly caused by high impedance interfaces that amplified the IF noise and EMG interference.V2 / V3 (cyan and orange) show a continuous downward and then upward curved baseline, with a ‘loose-recontact’ polarisation drift; V6 (green) shows multiple sharp jumps around 3500 s V6 (green) even shows several sharp jumps around 3500 s, which is a typical phenomenon of charge accumulation reaching amplifier saturation and then sudden discharge under high contact resistance.The difference in DC bias between limb leads was nearly 400 kCounts, indicating that the electrode potentials were out of balance between channels, and the amplifier had to sacrifice dynamic range to follow the large DC bias, further reducing the resolution of the true ECG signal.

And after using the electrode sheet:The baseline levelled off in all leads, with only V2 / V3 still drifting slowly and monotonically, but with a significantly reduced slope; V6 no longer showed any distortion jumps.The main ECG waveforms were well defined, with sharp R-peaks followed by S-valleys that could be used directly for threshold detection; the waveforms had smooth edges and were not nibbled by the 50 Hz sinusoidal disturbances.The longitudinal range is narrowed from 750 kCounts to 420 kCounts, proving that the amplifier is free from large potential bias and that more effective bits are allocated to the AC component.Taken together, the addition of metal electrodes significantly lowers contact impedance, reduces polarisation drift and amplifier saturation, and improves signal-to-noise ratio and detectability.

The figure above shows the ECG waveforms of the tester from the resting state to the exercise state. The P-QRS-T wave clusters in each lead are clearly discernible at rest, indicating stable electrode-skin impedance. Once exercise is initiated, there is an instantaneous synchronised spike in all leads, followed by sharp oscillations and baseline drift until stability is restored. This phenomenon is not a sudden change in the electrical activity of the heart itself, but is the result of motion artefacts superimposed on the ECG signal, which are triggered by the superimposition of the following three types of mechanisms:1. Electrode-skin microslip leading to sudden changes in polarisation potentials
When the wearer gets up and takes a step, the relative displacement of the skin and the microneedle electrodes produces a micron-level displacement, which destroys the original balanced electrical double layer; the polarisation potential can jump to hundreds of millivolts when it is instantly reconstructed, which is amplified by the front-end amplifier as a ‘large amplitude electrocardiogram’ all-channel amplification, forming synchronous spikes.2. Piezoelectric and friction effect of wire stretching and rebound.Although S-shaped wires can cushion large displacements, they are still stretched during one or two strides of sudden acceleration; the friction between the conducting fibres and the fabric, and the piezoelectric/friction charge redistribution of the fibres within the wire will inject equivalent noise along the entire wire, which will be manifested as a full-conductance common-mode jitter and DC bias jumps.3. Electromyography (EMG) coupled with acceleration noise When walking, the pectoral and intercostal muscles continue to contract, and the broadband EMG (20 Hz-300 Hz) spectrum generated by them falls within the 0.05-150 Hz bandpass range of the instrument; the micro-phonic effect of the cable generated by acceleration converts the low-frequency mechanical vibration into electrical signals, which will be reflected in the full-lead common mode jitter and DC bias jumps. The micro-phonic effect of the acceleration converts low-frequency mechanical vibrations into electrical signals, and the superposition of the two makes the waveform appear dense high-frequency burrs and baseline ripples.

Since the three disturbances occur almost simultaneously, the whole motion region is characterised by ‘ECG masked by full bandwidth noise superimposed on baseline drift and gradual recovery after saturation’. This can be mitigated in practice by the following means:Mechanical side: add local compression elastic bands at key leads on the chest to limit skin-electrode relative slip; add a TPU heat-sealed ‘limit bridge’ between the lead and the garment to reduce transient pulling. Electrode side: micro-cured hydrogel or biocellulose ring around the microneedle to increase the coefficient of friction and provide adhesion cushioning. Circuit side: Improve front-end ADC dynamic range and set up fast recovery anti-saturation circuitry; introduce tri-axial accelerometers at the MCU side and remove synchronous motion noise with adaptive filtering/template reconstruction algorithms. On the algorithmic side, motion artefacts are detected using full-lead co-modulation, and the RR intervals are automatically masked or interpolated with heart rate prediction to patch the RR intervals, ensuring that the downstream HRV or rhythm analyses are not skewed by extreme spikes.

In short, the dramatic fluctuations in the figure are not cardiac abnormalities, but are typical of multi-source mechanical perturbations of electrodes, wires, and amplifier links at the moment of transition from rest to motion; such motion artefacts can be significantly attenuated with improved fixation and signal processing algorithms for wearable ECGs, maintaining clinical-grade waveform quality.

\section{Conclusion}

In summary, this project takes the clinical 12-lead monitoring demand as the starting point, and completes the overall architecture design, prototyping and functional verification of the wearable ECG suit by focusing on the three key aspects of ‘electrode-conductor-signal chain’. The metal-coated microneedle electrodes prepared by DLP 3D printing and PVD process compress the interface impedance to the ideal range between the traditional dry electrode and wet electrode under the premise of ensuring the skin comfort; together with the S-shape conductive cloth routing and double hot-pressing encapsulation process, the conductor maintains the low impedance and mechanical integrity under the conditions of repeated bending, stretching and washing. Multi-scenario comparative experiments show that, after using microneedle electrodes and conductive paste, the resting ECG waveforms P, QRS, and T have a clear structure, improved signal-to-noise ratio, and significantly reduced baseline drift; the transition test from rest to motion also verifies the advantages of the S-shaped wires and improved fixation method in suppressing motion artifacts. Overall, the prototype not only meets the clinical monitoring requirements in terms of signal quality, but also lays the foundation for commercialisation with its flexible material system and mass production process. Future work will continue to optimise the surface structure of the microneedle and the details of the wire encapsulation, introduce low-power wireless modules and adaptive filtering algorithms, and realise round-the-clock and long-distance monitoring and big data analysis, so as to provide reliable support for personal health management and telemedicine.









\bibliographystyle{cas-model2-names}

\bibliography{cas-refs}



\end{document}